%% file: carres.tex
\documentclass[11pt,letterpaper]{article}
\usepackage[utf8]{inputenc}
\usepackage{amsmath}
\usepackage{amssymb}
\usepackage{amsfonts}
\usepackage{amsthm}
\usepackage{graphicx}
\usepackage{color}
\usepackage{hyperref}
\usepackage{cite}
\usepackage{tikz}
\usepackage{xspace}
\usepackage{needspace}
\usepackage[section]{placeins}
\usepackage[margin=1in]{geometry}
\usetikzlibrary{decorations.pathmorphing}
\newtheorem{theorem}{Theorem}[section]

\newtheorem{lemma}[theorem]{Lemma}
\newtheorem{corollary}[theorem]{Corollary}

\theoremstyle{definition}
\newtheorem{definition}[theorem]{Definition}

\newtheorem*{remark}{Remark}
\newtheorem{openproblem}{Open Problem}

\makeatletter
\newtheorem*{rep@theorem}{\rep@title}
\newcommand{\newreptheorem}[2]{%
\newenvironment{rep#1}[1]{%
 \def\rep@title{#2 \ref{##1}}%
 \begin{rep@theorem}}%
 {\end{rep@theorem}}}
\makeatother

\newreptheorem{theorem}{Theorem}

\newtheorem{algorithm}[theorem]{Algorithm}
\let\oldalgorithm\algorithm
\renewcommand{\algorithm}{\oldalgorithm\normalfont}

\newcounter{choice}

\newcommand\ignore[1]{}

\newcommand{\termasm}[1]{\mathcal{A}_{\Box}[{#1}]}
\newcommand{\prodasm}[1]{\mathcal{A}[{#1}]}
\newcommand{\dom}[1]{{\rm dom}(#1)}

\newcommand{\north}{\mathrm{N}}
\newcommand{\south}{\mathrm{S}}
\newcommand{\east}{\mathrm{E}}
\newcommand{\west}{\mathrm{W}}

\newcommand{\vect}{\overrightarrow}
\newcommand{\resp}{respectively\xspace}

\title{The self-assembly of paths and squares at temperature 1}
\author{Pierre-\'Etienne Meunier
\thanks{Computer Science, California Institute of Technology. \protect\url{pmeunier@caltech.edu}.
Supported in part by National Science Foundation Grant CCF-1219274.}}
\date{}
\begin{document}

\maketitle

\begin{abstract}

We prove that the number of tile types required to build squares of
size $n\times n$, in Winfree's abstract Tile Assembly Model, when
restricted to using only non-cooperative tile bindings, is at least
$2n-1$, which is also the best known upper bound. Non-cooperative
self-assembly, also known as ``temperature 1'', is where tiles bind to
each other if they match on one or more sides, whereas in cooperative
binding, some tiles can bind only if they match on multiple sides.

Our proof introduces a new programming technique for temperature 1,
that disproves the very intuitive and commonly held belief that, in
the same model, assembling \emph{paths} between two points $A$ and $B$
cannot be done with less tile types than the Manhattan distance
$\|\overrightarrow{AB}\|_1$ between them. Then, we prove a necessary
condition for these ``efficient paths'' to be assembled, and show that
this necessary condition cannot hold in completely filled squares.

This result proves the oldest conjecture in algorithmic self-assembly,
published by Rothemund and Winfree in STOC 2000, in the case where
growth starts from a corner of the square. As a corollary, we
establish $n$ as a lower bound on the tile complexity of the general
case. The problem of determining the minimal number of tile types to
self-assemble a shape is known to be $\Sigma_2^p$-complete.

\end{abstract}

\section{Introduction}
\label{introduction}

\input{intro}

\subsection{Main results}

Although a number of terms have not been formally defined, we give an
overview of our two main results now. See the definitions in section
\ref{definitions}. Our ultimate goal is to prove Rothemund and
Winfree's conjecture \cite{RotWin00} that a tile assembly system
assembling only squares of size $n\times n$, from a seed of size $1$,
has at least $2n-1$ tile types. Our first result disproves a statement
stronger than this conjecture; namely, that the tile complexity of a
square is the same as the tile complexity of its diagonal. Although
widely believed, this statement is false:

\begin{reptheorem}{efficient paths}
Let $n$ be an integer.  There is a tile assembly system
$\mathcal{T}_n=(T_n,\sigma_n,1)$, and two points $A,B\in\mathbb{Z}^2$, such
that $\|\vect{AB}\|_1=n$, the terminal assemblies of $\mathcal{T}_n$
are all finite, they all include a path from $A$ to $B$, and $|T_n|=
4n/5+O(1)$.
\end{reptheorem}

The fact that the constructions of this theorem are possible, even
though they are quite elementary, is not obvious at all; indeed, the
intuition from words and automata theory is that any attempt to
``reuse'' tile types will enable us to ``pump'' the path, as in the
pumping lemma of finite automata \cite{Sipser}, and thus any tile
assembly system that can produce paths repeating tile types will also
be able to produce ultimately periodic, infinite paths. This intuition
is valid in a restricted setting where two adjacent tiles always agree
on their abutting sides \cite{Manuch-2010,Doty-2011}.

Then, we will prove the following theorem, which gives the optimal
lower bound on the tile complexity of squares, when growth starts from
a corner:

\begin{reptheorem}{inacorner}
\label{thm:squares}
Let $\mathcal{T}=(T,\sigma,1)$ be a temperature 1 tile assembly
system, with $\sigma$ a single tile at $(0,0)$, and $n$ an integer. If
all terminal assemblies producible by $\mathcal{T}$ are of domain
$\{0,\ldots,n-1\}^2$, then $|T|\geq 2n-1$.
\end{reptheorem}

\section{Definition and preliminaries}
\label{definitions}

We begin by defining the two-dimensional abstract tile assembly
model. A \emph{tile type} is a unit square with four sides,
each consisting of a glue \emph{label} and a nonnegative integer
\emph{strength}. We call a tile's sides north, east, south, and west,
respectively, according to the following picture:
\begin{center}
\begin{tikzpicture}
\draw(0,0)rectangle(1,1);
\draw(0,0.5)node[anchor=east]{West};
\draw(1,0.5)node[anchor=west]{East};
\draw(0.5,0)node[anchor=north]{South};
\draw(0.5,1)node[anchor=south]{North};
\end{tikzpicture}
\end{center}
We assume a finite set $T$ of tile types,
but an infinite supply of copies of each type. An \emph{assembly}
is a positioning of the tiles on the discrete plane $\mathbb{Z}^2$,
that is, a partial function $\alpha:\mathbb{Z}^2\dashrightarrow T$.

We say that two tiles in an assembly \emph{interact}, or are
\emph{stably attached}, if the glue labels on their abutting side are
equal, and have positive strength.  An assembly $\alpha$ induces a
weighted \emph{binding graph} $G=(V,E)$, where $V=\dom{\alpha}$, and
there is an edge $(a,b)\in E$ if and only if $a$ and $b$ interact, and
this edge is weighted by the glue strength of that interaction.  The
assembly is said to be $\tau$-stable if any cut of $G$ has weight at
least $\tau$.

A \emph{tile assembly system} is a triple $\mathcal{T}=(T,\sigma,\tau)$,
where $T$ is a finite tile set, $\sigma$ is called the \emph{seed}, and
$\tau$ is the \emph{temperature}. Throughout this paper, we will always
have $|\dom{\sigma}|=1$ and $\tau=1$. Therefore, we can make the
simplifying assumption that all glues have strength one without
changing the behavior of the model.

Given two assemblies $\alpha$ and $\beta$, we write
$\alpha\rightarrow_1^{\mathcal{T}}\beta$ if we can get $\beta$ from $\alpha$ by
the binding of a single tile, $\dom{\alpha}\subseteq\dom{\beta}$, and
$|\dom{\beta}\setminus\dom{\alpha}|=1$.  We say that $\gamma$ is
\emph{producible} from $\alpha$, and write
$\alpha\rightarrow^{\mathcal{T}}\gamma$ if there is a (possibly empty)
sequence $\alpha=\alpha_1,\ldots,\alpha_n=\beta$ such that
$\alpha_1\rightarrow_1^{\mathcal{T}}\ldots\rightarrow_1^{\mathcal{T}}\alpha_n$.

A sequence of $k\in\mathbb{Z}^+ \cup \{\infty\}$ assemblies
$\alpha_0,\alpha_1,\ldots$ over $\mathcal{A}^T$ is a
\emph{$\mathcal{T}$-assembly sequence} if, for all $1 \leq i < k$,
$\alpha_{i-1} \to_1^\mathcal{T} \alpha_{i}$.

The \emph{productions} of a tile assembly system $\mathcal{T}=(T,\sigma,\tau)$,
written $\prodasm{\mathcal{T}}$, is the set of all assemblies producible
from $\sigma$. An assembly $\alpha$ is called \emph{terminal} if there
is no $\beta$ such that $\alpha\rightarrow_1^{\mathcal{T}}\beta$. The
set of terminal assemblies is written $\termasm{\mathcal{T}}$.

An important fact about temperature 1 tile assembly, that we will use
heavily, is that any path of the binding graph can grow immediately
from the seed, independently from anything else. Formally, a
\emph{path} $P$ is a sequence of tile types along with positions, that
is, of elements of $T\times\mathbb{Z}^2$, such that no position occurs
more than once in $P$, and for all $i$, the positions of $P_i$ and $P_{i+1}$
are adjacent in the lattice grid of $\mathbb{Z}^2$.

For any path $P$ and integer $i$, we write $(x_{P_i},y_{P_i})$ the
coordinates of $P_i$'s position.  Moreover, if $i<n$, we say that the
\emph{output side} of $P_i$ is the side adjacent to $P_{i+1}$, and if
$i>1$, that its \emph{input side} is the side adjacent to
$P_{i-1}$. This means that the first tile of a path does not have an
input side, and the last one does not have an output side.  Remark
that this definition of input/output sides is only relative to a
\emph{path}, and not to the tiles themselves; indeed, the tiles,
including the first and last ones, may have other glues, not used by
the path.

Also, for any path $P=P_0,P_1,\ldots,P_{|P|-1}$, and any integer $i$,
we call the \emph{right-hand side} (\resp \emph{left-hand side}) of
$P_i$ the side of that is between its input and output sides, in
counterclockwise (\resp clockwise) order. When there is no ambiguity,
we will say ``the right-hand side (\resp left-hand side)'' of $P$
itself to mean ``the right-hand side (\resp left-hand side) of some
tile of $P$''.

Finally, for $A,B\in\mathbb{Z}^2$ (\resp for $A,B\in
T\times\mathbb{Z}^2$), we use the notation $\vect{AB}$ to mean ``the
vector from $A$ to $B$'' (\resp from the \emph{position} of $A$ to the
\emph{position} of $B$), and the \emph{Manhattan distance} between
$A=(x_A,y_A)$ and $B=(x_B,y_B)$, written $\|\vect{AB}\|_1$, is
$|x_A-x_B|+|y_A-y_B|$. We also call $O$ the \emph{origin} on
$\mathbb{Z}^2$, i.e. the point of coordinates $(0,0)$.

\needspace{7\baselineskip}\subsection{The known upper bound}

\begin{figure}[ht]
\begin{center}
\includegraphics[scale=5]{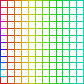}
\end{center}
\caption{The comb design for size $10\times 10$}
\label{comb}
\end{figure}

The only known way to assemble squares of size $n\times n$ at
temperature 1 with $2n-1$ tile types, is by using the ``comb'' design
of Figure \ref{comb}, already described in Rothemund and Winfree's
paper \cite{RotWin00}.

\section{Building efficient paths}

A major obstacle in proving the claimed lower bound for squares, is
that building only a square's diagonal can require less tile types than
building the whole square. In this section, we show the following
result:

\begin{theorem}
\label{efficient paths}
There is a tile assembly system $\mathcal{T}=(T,\sigma,1)$, such that
for all terminal assembly $a\in\termasm{\mathcal{T}}$, $a$ is finite and
of width $\frac{5(|T|+2)}{4}-23$.

\begin{proof}
Since a path of height $n$ that is monotonic in the y-dimension, and has
less than $n$ tile types with input side south, can be ``pumped'',
our path will need to have ``caves'', or non-monotonic subpaths, and
reuse them several times. Moreover, in order for all the assemblies of
$\mathcal{T}$ to be finite, the caves must be exited by a different
path every time; however, since we want the caves to ``save'' tile
types, these ``exit paths'' cannot all be new tile types.  Therefore,
one possible way to solve these constraints is to grow a regular
monotonic path $P_0$ first, then build a cave $C$, and reuse a part of
$P_0$ as its exit path. The next time we want to reuse $C$, we can use
another part of $P_0$ as its exit path.

If the exit paths used in previous instance of a cave are all blocked
in new instances, we will get only finite assemblies.
Figure \ref{figure:efficient path 1} is an example of such a path.

\begin{figure}[ht]
\begin{center}
\includegraphics[scale=22]{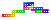}
\end{center}
\caption{Reusing caves and exit paths}
\label{figure:efficient path 1}
\end{figure}

This figure, however, is not a terminal assembly; since our caves, and
their exit paths, are used several times, the same assemblies can grow
from all their repetitions. Fortunately, we can arrange the shape of
our main path so that no collision ever happens between these
repetitions, like on Figure~\ref{figure:efficient path 2}.
Now, this figure has 38 tile types, and is of width 27; it does not
yet save tile types. But by inserting:
\begin{itemize}
\item $n$ new tile types in place of glue 6,
\item $n$ new tile types in place of glue 14,
\item $n$ new tile types in place of glue 24, and
\item $n$ new tile types in place of glue 26
\end{itemize}

Zooming in may be needed to read these numbers on Figures
\ref{figure:efficient path 1} and \ref{figure:efficient path
  2}. Printable versions are included, in Appendices
\ref{printable-eff1} and \ref{printable-eff2}.

We add only $4n$ tile types, but the assembly is now $5n$ wider, and
the result follows. An example of path that actually saves tile types
is given on Figure \ref{figure:efficient path}.

\begin{figure}[!ht]
\begin{center}
\includegraphics[scale=22]{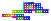}
\end{center}
\caption{Early branches, not blocking the main assembly}
\label{figure:efficient path 2}
\end{figure}

\begin{figure}[!ht]
\begin{center}
\includegraphics[scale=5]{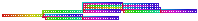}
\end{center}
\caption{A path of width 112 with 106 tile types}
\label{figure:efficient path}
\end{figure}
\end{proof}
\end{theorem}

\section{Building filled squares requires \texorpdfstring{$2n-1$}{2n-1} tile types}

In this section, we prove that if a tileset has less than $2n-1$ tile
types and all its terminal assemblies are of domain
$\{0,1,\ldots,n-1\}^2$ then one of its productions is a path that does
not ``save'' tile types. Our technique to prove this will be the
following: assuming we are given such a tileset, we will first choose,
using Algorithm \ref{path building}, the assembly sequences that
``lose'' as much information as possible about their past, so as to
``confuse'' the tileset; then, if these assembly sequences can still
build efficient paths, we will prove a necessary condition on these
(this is Lemma \ref{lem:tentacular}), that cannot hold in completely
filled squares (Theorem \ref {thm:inacorner}).

\subsection{A path-building algorithm}

We first define the algorithm we use to find assembly sequences that
suit our purposes.
There are two possible ``priority modes'' for this algorithm, namely
\emph{right-priority} and \emph{left-priority}:

\begin{algorithm}

\label{path building}
Let $\mathcal{T}=(T,\sigma,1)$ be a tile assembly system, $\preceq$ be
an ordering on $T$, and $S$ be a non-empty subset of the set of paths
producible by $\mathcal{T}$. Let $P_0$ be the initial path, with just
$\sigma$. For any $i\geq 0$, if several tiles can bind to $P_i$, let
$P_{i+1}$ be the one such that:

\begin{enumerate}
\item If $P_{i}$ is of the same type as a previous tile $P_h$ on $P$,
  with $h<i$, whose output side $s$ is distinct from $P_i$'s input
  side, then grow $P_{h+1}$, on side $s$ of $P_{i}$.
  \label{case:rep}
\item Else, if it is possible, let $P_{i+1}$ be the tile such that:
\begin{enumerate}
\item $P_{i+1}$ binds to the first side of $P_i$ from its input side, in
  clockwise order if we are building a left-priority path, and in
  counterclockwise order if we are building a right-priority one.

\item There is at least one path in $S$, of which $P_0,P_1,\ldots,P_{i+1}$
  is a prefix.

\item If there are several such choices, we choose the smallest tile
  with respect to $\preceq$.\setcounter{choice}{\value{enumi}}\label{choice}
\end{enumerate}

\item Finally, if none of the previous cases is possible to follow (for
  instance, because following case \ref{case:rep} resulted in a
  collision with a previous part of the assembly), but a new branch
  can still grow from the existing assembly, and produce a path of
  $S$, then follow it. Else, the algorithm halts.

  We call the last tile that was placed before the present case the
  \emph{collision tile}.
  In the special case that the current assembly before applying this
  case is already in $S$, we will adopt the convention that the \emph{empty
    branch} can start.
  \label{algo:col}
\end{enumerate}
\end{algorithm}

Remark that this algorithm does not guarantee that the final path will
be in $S$.  If all producible paths end up being paths of $S$, it will
be the case. But else, it means that we can ``prevent'' paths of $S$
from growing. We will use this property heavily in the rest of our
proof: $S$ will be the set of all paths that reach some point in the
square, and by definition, all the points in a \emph{completely
  filled} square must be covered.

\subsection{Building filled squares from a seed in a corner}
\label{inacorner}
We begin by showing that a ``path editing'' operation is possible on
the paths, built by Algorithm \ref{path building}, that save tile types.

\begin{definition}
  \label{tentacular paths}
  A \emph{left-tentacular} (\resp \emph{right-tentacular}) path is a
  path $P$ such that the following conditions all hold:
  \begin{itemize}
    \item at least two tiles of $P$ are of the same type.
      Let $i$ and $j$ be the indices of two tiles of $P$ of the same type.
    \item $P$ forked after a collision in case
      \ref{algo:col} of Algorithm \ref{path building}, at some position
      $k>j$, and this branch started on the left-hand side (\resp right-hand
      side) of $P_k$.
    \item This same branch can also start and completely grow from
      tile $P_{k-j+i}$, after $P$ is itself completely grown.
  \end{itemize}

  Moreover, we call $\vect{P_jP_i}$ a \emph{contraction vector} of
  $P$, and \emph{tentacles} the early restarts of $P_{k,k+1,\ldots |P|-1}$.

  We say that a path is \emph{two-way tentacular} if it is both left-
  and right-tentacular.

\end{definition}

\begin{figure}[ht]
\begin{center}
  \includegraphics[scale=3]{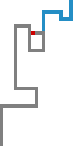}\hspace{1cm}
  \input{tentac}
\end{center}
\caption{A right-tentacular path, with the branch in blue, and the collision tile in red.
A contraction vector is drawn on the right-hand drawing.}
\label{fig:tentac}
\end{figure}

\begin{definition}

  \label{fragile}
  Let $\mathcal{T}=(T,\sigma,1)$ be a tile assembly system. A path
  $P\in\prodasm{\mathcal{T}}$ is said to be \emph{fragile} if:

  \begin{itemize}
    \item a tile type is repeated on $P$, at positions $i$ and $j$
      ($i<j$), and $P_{j,j+1,\ldots,|P|-1}$, translated by
      $\vect{P_jP_i}$, can also start growing from $P$ immediately
      after $i$.
    \item and $P_{i,i+1,\ldots,|P|-1}$ cannot grow completely from the
      resulting assembly.
  \end{itemize}

  In the rest of the proof, we will call \emph{breaking} $P$, the
  choice to start growing $P_{j,j+1,\ldots|P|-1}$, translated by
  $\vect{P_jP_i}$, immediately after $i$ before growing
  $P_{i,i+1,\ldots,|P|-1}$. Moreover, the smallest suffix of $P$ that
  can branched earlier than its original first point on $P$ is called
  the \emph{breaking branch}.

\end{definition}

See Figure \ref{fig:visible} for an example of a fragile path (at this
point, though, the colors of that figure are not yet defined).

\begin{lemma}
  \label{lem:tentacular}
  Let $A=(x_A,y_A)$ be a point of $\mathbb{Z}^2$, and
  $\mathcal{T}=(T,\sigma,1)$ be a tile assembly system such that the
  following conditions all hold:
  \begin{itemize}
    \item $\dom{\sigma}=\{O\}$.
    \item All assemblies can reach $A$, that is,
      $\forall \alpha\in\termasm{\mathcal{T}}, A\in\dom{\alpha}$.
    \item There is a path $P\in\prodasm{\mathcal{T}}$ from $O$ to $A$,
      built using Algorithm \ref{path building} with parameter $S$ of
      the algorithm being the set of all producible paths of
      $\mathcal{T}$ from $O$ to $A$, and $P$ has strictly less than
      $\|\vect{OA}\|_1-1$ tile types.
    \item $P$ is such that for all $i$, $y_{P_i}\in\{0,\ldots,y_A\}$.
  \end{itemize}

  Then $P$ is either fragile or two-way tentacular.

  \begin{proof}

    Let $P$ be a path from $O$ to $A$, built using Algorithm \ref{path
      building}. Without loss of generality, we assume that
    $y_A\geq 0$ and $x_A\geq 0$ (we get other cases by rotating or flipping
    the argument).

    We first introduce the idea of \emph{visible glues}: we say that a
    glue between two tiles of $P$ is visible from the east (\resp from
    the south) if $P_i$ interacts with $P_{i+1}$ on its north face
    (\resp on its east face), and the glue between them is the
    rightmost (\resp lowest) one, on an infinite horizontal (\resp
    vertical) line between $P_i$ and $P_{i+1}$. Moreover, we adopt the
    convention that the rightmost tile on row $y=y_A$ has its north
    glue visible from the east, and its east glue visible from the
    south (even if this tile has 0-strength glues on these sides).

    \begin{figure}[ht]
      \begin{center}
        \includegraphics[scale=3]{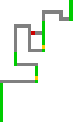}\hspace{1cm}
        \includegraphics[scale=3]{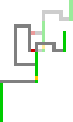}
      \end{center}
      \caption{Tiles with glues visible from the east are in green or
        yellow; yellow tiles are of the same type, and red tiles are
        collisions. In this case, $P$ is fragile.}
      \label{fig:visible}
    \end{figure}

    Let us call $P_{\east}$ the tiles of $P$ that have their north
    glue visible from the east, and $P_{\south}$ the tiles of $P$ that
    have their east glue visible from the south.  First, there are at
    least $y_A$ tiles in $P_{\east}$, and $x_A$ tiles in $P_{\south}$.
    We prove this now for $P_\south$:
    indeed, for all $x\in\{0,\ldots,x_A-1\}$, the visible tile of $P$
    on column $x$ has output side east. To see this, suppose, for
    the sake of contradiction, that some tile $P_i$ is visible from
    the south, and has output side $\west$. Then, draw an infinite
    horizontal half-line from the east side of $P_i$ to the
    south. Along with $P_{0,\ldots,i}$ and a horizontal infinite line
    at $y=0$, it partitions the plane into two connected components,
    by Jordan's curve theorem (see Figure \ref{fig:jordan}).

    Therefore, because $P$ does not intersect itself, and for all $i$,
    $y_{P_i}\geq 0$, it must necessarily cross the vertical line at
    the east of $P_i$ again before reaching $A$. Thus, $P_i$ cannot be
    visible from the south, since $P$ has at least one tile
    below $P_i$.

    \begin{figure}[ht]
    \begin{center}
      \input{jordan1}
    \end{center}
    \caption{Jordan's curve theorem in action, with the enclosed region in blue}
    \label{fig:jordan}
    \end{figure}
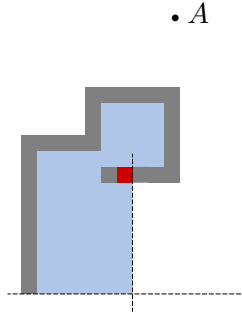

    The same argument, rotated by $\pi/2$, shows that $P_{\east}$ has
    at least $y_A$ tiles.
    Furthermore, except possibly for the last tile of $P$,
    $P_{\east}$ and $P_{\south}$ are disjoint, because each tile has
    at most one output side:
    $P_\east\cap P_\south\subseteq\{P_{|P|-1}\}$, and therefore $|P_\east\cup P_\south|\geq \|\vect{OA}\|_1-1$.

    Now, because $P$ was grown using Algorithm \ref{path building}, if
    a tile type is repeated in $P_\east$, say at positions $P_i$ and
    $P_j$, with $i<j$, then the translation of $P_{i,i+1,\ldots,j}$ by
    $\vect{P_iP_j}$ started growing immediately after $P_j$, by case
    \ref{case:rep} of Algorithm \ref{path building}, ultimately
    crashing into something, possibly after several repetitions
    (because we assumed that all productions of $\mathcal{T}$ are
    finite). We call this collision ``collision $C_0$''. The next step
    of the algorithm after this happened was thus necessarily by case
    \ref{algo:col} (because only that case handles collisions), and a
    new (possibly empty) branch was started from the existing
    assembly, and reached $A$ (because of our hypothesis that all
    terminal assemblies have $A$ in their domain).

    Therefore, an early restart of this branch can also grow from
    $P_{i,i+1,\ldots,j}$. There are two cases:

    \begin{enumerate}
    \item If it can grow completely from $P$, it means that $P$
      is right-tentacular,  by definition of right-tentacular.

    \item Else, this new branch crashes into something. We call this
      crash ``collision $C_1$''. There are two cases, according to
      where it crashes:
      \begin{enumerate}
        \item Either it crashes into a part of $P$ grown before $P_i$.
          In this case, first observe that the north glue of $P_i$ is
          still visible from the east after this operation: indeed,
          the north glue of $P_j$ was visible from the east before, so
          no $P_k$, for $k>j$, is on the right of $P_j$ on a
          horizontal line at $y=y_{P_j}$. And since we only translated
          these tiles by $\vect{P_jP_i}$, these translated tiles are
          not on the right of $P_i$ on a horizontal line at
          $y=y_{P_i}$ either (see for instance Figure \ref{fig:visible}).

          We now prove that $P$ is fragile. Indeed, grow it until
          index $i$. Then, grow a translation by $\vect{P_jP_i}$ of
          $P_{j,j+1,\ldots,|P|-1}$, as long as it can possibly
          grow. We call $R$ the resulting subpath (the longest prefix
          of the translation of $P_{j,j+1,\ldots,|P|-1}$ that can
          grow). We claim that the original $P$ cannot grow anymore
          from this new assembly $Q=P\cup R$, and this is the
          definition of $P$ being fragile.

          This is because of the definition of $P_\east$ and
          $P_\south$: we first argue that collision $C_1$ encloses a
          portion of the plane: this is indeed an application of
          Jordan's curve theorem, because a collision between two
          connected paths closes a curve in the plane.  But then,
          since $P_i$ is still visible from the east or from the south
          in $Q$, and the side of $P_i$ exposed to the south or the
          east is its right-hand side (this follows from the fact that
          $P_i$'s output side is its north side if visible from the
          east, and its east side if visible from the south, and the
          definition of the left- and right-hand sides of a tile on a
          path), the breaking branch starts from some $P_k$ (for
          $k>j$), and branches from its right-hand side.

          Therefore, if we grow this branch early, at position
          $k-j+i$, and it crashes into a previous part of the
          assembly, then by Jordan's curve theorem, we enclose the
          left-hand sides of its tiles; but since this branch forks
          from the right-hand side of $P$, the enclosed region is the
          region in which $P$ grew (because $P$ is then a ``left
          turn'' from $P_{0,1,\ldots,i}\cup R$). Therefore, since for
          all $i$, $y_{P_i}\in\{0,\ldots,y_A\}$, $R$ cannot grow
          strictly higher than $y_A$; thus, $P$ cannot reach $A$
          anymore from this assembly, which means that it cannot grow
          to its original endpoint, and therefore, it is fragile.

        \item Or it crashes into a part of $P$ grown after $P_i$, in
          which case we also choose to grow it before that part. Since
          $P$ is a \emph{sequence} of points, it will not be able to
          grow completely after that, since a tile will already be
          present next to the collision tile of collision $C_1$.
      \end{enumerate}
    \end{enumerate}

    Finally, the same argument on the tiles of $P_\west$ and $P_{\north}$
    (the tiles with their north glue visible from the left, and their
    east glue visible from the north, respectively),
    proves that $P$ is also left-tentacular or fragile. Remark that
    $P_\west\cup P_\north$ is not necessarily disjoint from
    $P_\east\cup P_\south$ (in the case where $P$ has at most
    $2n-1$ tile types, these two sets may even be equal).
  \end{proof}
\end{lemma}

\begin{remark}
Consider the tentacles of Figure \ref{figure:efficient path 2},
restarted from the path of Figure \ref{figure:efficient path 1}:
The right tentacles are after glue 32, and then glue 36, and
the left one after glue 35.
\end{remark}

\needspace{3\baselineskip}We can now prove the claimed result:
\begin{theorem}
  \label{thm:inacorner}
  Let $\mathcal{T}=(T,\sigma,1)$ be a tile assembly system whose
  terminal assemblies' domains are all $\{0,\ldots,n-1\}^2$, and
  such that $\dom{\sigma}=\{(0,0)\}$.
  Then:

  $$|T|\geq 2n-1$$

  \begin{proof}

    Let $A$, $B$, $C$ and $D$ be the following points of a square that
    $\mathcal{T}$ can assemble:
    \begin{center}
    \begin{tikzpicture}[scale=1.5]
      \draw(0,1)node[shape=coordinate](a){};
      \draw(1,1)node[shape=coordinate](b){};
      \draw(1,0)node[shape=coordinate](c){};
      \draw(0,0)node[shape=coordinate](d){};
      \draw(0.234,0.45)node[shape=coordinate](o){};
      \draw(a)node[anchor=south east]{$A$} --
      (b) node[anchor=south west]{$B$} --
      (c) node[anchor=north west]{$C$} --
      (d) node[anchor=north east]{$D=O$} -- cycle;
    \end{tikzpicture}
    \end{center}

    Since $\mathcal{T}$ can fill this square, it must contain in
    particular a path $P^0$ from $O$ to $A$, that we can build
    in the right-priority mode of Algorithm \ref{path
      building}.

    We define a sequence $(S_i)_i$ of assemblies, $S_0$ being the
    assembly where only $P^0$ has grown.  For any $i\geq 0$, let $A_i$
    be the leftmost point on the right-hand side of $S_i$, and not in
    $S_i$. If there are several such points, let $A_i$ be the highest
    one. Now, let $P^{i+1}$ be a right-priority path from $O$ to
    $A_i$, built using Algorithm \ref{path building}. Moreover, let
    $Q^{i+1}$ be the prefix of $P^{i+1}$ that stops at the last
    occurrence of $P^{i+1}$'s highest point.

    There are three main cases:
    \begin{enumerate}
    \item If $Q^{i+1}$ is fragile, we break it. Let $R^{i+1}$ be the
      assembly resulting from that operation. In the case where
      $Q_{i+1}$ shares parts with some other paths of $S_i$, this
      operation may also break these paths.  Let $S_{i+1}$ be the
      assembly containing $R^{i+1}$ and all the parts of $S_i$ that
      can regrow from it.
      \label{case:fragile}
    \item If $Q^{i+1}$ is right-tentacular, there are two subcases:
      \begin{enumerate}
      \item
        One of its contraction vectors $\vect{v_{i+1}}$ is to the
        left, i.e. $x_{\vect{v_{i+1}}}\leq 0$. In this case, we can do
        the same as in case \ref{case:fragile} above. Indeed, since
        the last point of $Q^{i+1}$ is next to the leftmost point on
        the right of $S_i$, contracting it to the left will
        necessarily end in a collision with $S_i$, before the end of
        $Q^{i+1}$.

        Indeed, this is clear in the case of $Q^{i+1}=P^{i+1}$, i.e. $A_{i}$
        is the highest point of $P^{i+1}$, as well as in the case where
        $A_i$ is on the left of $Q^{i+1}$.

        Else, $P^{i+1}$ is longer than $Q^{i+1}$, and the subpath
        $P^{i+1}\setminus Q^{i+1}$ grows after $Q^{i+1}$, by
        definition of $Q^{i+1}$ being a prefix of $P^{i+1}$.
        Therefore, $A_i$ is on the right of
        $Q^{i+1}$: indeed, by Jordan's curve theorem, on the closed
        curve pictured on Figure \ref{fig:jordan2} (the curve defined
        by $Q^{i+1}$, two horizontal lines, immediately above and
        below $Q^{i+1}$, and a vertical line on the right of the
        square), $P^{i+1}\setminus Q^{i+1}$ is \emph{inside} a closed
        region of the plane, including a point on the right of
        $Q^{i+1}$ (for instance the first point of $Q^{i+1}$).

        \begin{figure}[ht]
          \begin{center}
            \input{jordan2}
          \end{center}
          \caption{Jordan's curve theorem.
            The enclosed region is colored blue, $Q^{i+1}$ is in grey,
            and $P^{i+1}\setminus Q^{i+1}$ is in yellow.}
          \label{fig:jordan2}
        \end{figure}
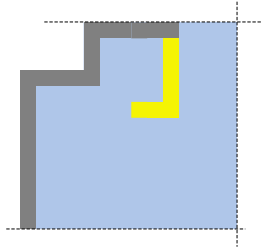

        Therefore, $Q^i$ is necessarily already in
        $S_i$, because $A_i$ was chosen as the leftmost point on the
        right of $S_i$.  Thus, there is an $h<i$ such that $A_h$ is
        the highest (and last) point of $Q^{i+1}$, and we can apply
        the same argument as in the case where $Q^{i+1}=P^{i+1}$,
        proving that $P^{i+1}$ is fragile.
        \label{right-tentacular-right}

      \item Else, all its contraction vectors are strictly to the
        right.  It is not possible that $B\in\dom{P^{i+1}}$, for else
        $B$ would also be in $\dom{Q^{i+1}}$ (indeed, by definition of
        $Q^{i+1}$, it includes the last occurrence of the highest
        point on $P^{i+1}$) and then one of the tentacles, when grown
        alone, would grow out of $\{0,\ldots,n-1\}^2$, contrarily to
        our hypothesis that all the productions of $\mathcal{T}$ stay within
        $\{0,\ldots,n-1\}^2$.  Therefore, in this case, we let the
        tentacle grow as long as it can, until it reaches its highest
        point or else crashes into something.

        We continue this process with $S_{i+1}$ being the union of
        $S_i$, $P^{i+1}$, and the longest prefix of the tentacles that
        can grow).
        \label{right-tentacular-left}
      \end{enumerate}

    \item Else, by Lemma \ref{lem:tentacular}, $Q_{i+1}$ has at least
      $\|OA_i\|_1-1$ tile types. There are two subcases:
      \begin{enumerate}
      \item if $A_i\neq B$, we let $S_{i+1}=S_i\cup P_{i+1}$, and
        resume the construction.
      \item if $A_i=B$, the construction is over. Indeed, in this case
        $P^{i+1}=Q^{i+1}$, and $P^{i+1}$ is a path from $O$ to $B$,
        with at least $\|\vect{OB}\|_1-1=2n-1$ distinct tile types.
      \end{enumerate}
    \end{enumerate}

    In order to conclude the proof, we need to argue that this
    construction halts, even though cases \ref{case:fragile}
    and \ref{right-tentacular-right} seem to make the assembly
    smaller. First, all the paths we grow in this construction are
    right-priority paths. Therefore, after these two cases happen, the
    same points $A_i$ will appear again as $A_j$, for $j>i$, but in
    this case, the path that we can build from $O$ to $A_j$ will
    necessarily be ``less right-priority'' than the original ones,
    meaning that they will turn left earlier than $P^i$. Indeed, in
    both case \ref{case:fragile} and \ref{right-tentacular-right},
    we break $P^i$, and add the resulting assembly to $S_i$.

    To justify this implication, we need to examine what can possibly
    happen to these ``broken paths'' in greater detail.  When a
    path is broken, and we branch and grow a new one to reach $A_j$,
    one of two things could happen:
    \begin{enumerate}
    \item Either $P^j$ is fragile or right-tentacular with a
      contraction vector to the left, in which case nothing can happen
      to earlier broken parts. The assembly sequence that we use to
      prove this is the following: First grow all the parts that can
      grow on the left of $P^j$, then the broken $P^j$. Then, grow the
      earlier broken parts on its right. Either these parts are on the
      left of $P^j$'s breaking branch, in which case they are also
      enclosed, or they are on its right, but in this case, even if a
      collision happens between the breaking branch of $P^j$, and
      these parts, this collision is either with the original fragile
      path -- which still keeps it broken -- or with an earlier
      breaking branch -- which still keeps the original path enclosed,
      and therefore broken.

      \begin{remark}
        A fundamental thing about this process, that we are using
        here, is that breakings always happen between a breaking
        branch \emph{on the right of the path}, and the path itself.

        This argument would fail in the case where collisions between
        $P^j$ and an earlier broken part could ``open'' an earlier
        enclosed part, that would ``free'' a formerly broken
        path. Indeed, because of the very fact that these parts are
        broken, there is always the possibility to regrow a breaking
        branch until re-breaking the path.
      \end{remark}

    \item Else, the construction will continue. If $P^j$'s eventual
      tentacles crash into something, then this crash does not affect
      any of the arguments, since we only let them grow until their
      first collision.
    \end{enumerate}
  \end{proof}
\end{theorem}

\begin{corollary}
\label{cor}
Let $\mathcal{T}=(T,\sigma,1)$ be a tile assembly system whose terminal
assemblies are all of domain $\{0,\ldots,n-1\}^2$, and such that
$|\dom{\sigma}|=1$. Then $|T|\geq n$.
\begin{proof}

Let $\Sigma$ be the position of the seed.  We can use the technique of
Theorem \ref{inacorner} on all the rectangles with diagonals $(\Sigma
A)$, $(\Sigma B)$, $(\Sigma C)$ and $(\Sigma D)$. At least one is of
width and height at least $n/2$, and verify the assumptions of Lemma
\ref{lem:tentacular}, and the proof of Theorem \ref{thm:inacorner} can
be applied, either directly, or by ``turning in the other direction'',
i.e. looking for a non-left-tentacular path.
\end{proof}
\end{corollary}

\section{Future work}

The next step, in proving the fully general conjecture, is to extend
Lemma \ref{lem:tentacular} to the case where the seed can be
anywhere in the square. The reason why it does not apply to that case,
is that a ``lower restart'' is not necessarily an ``early restart'';
indeed, let $P$ be a path from $O$ to $D$, and $Q$ be a path from $O$ to $B$.
If $P_i$ and $Q_j$ are of the same type for some $i$ and $j$, then restarting
$P_{i,i+1,\ldots,|P|-1}$ from $Q_j$ does not result in a competition for growth between this
``lower branch'' and $Q$, since the tiles of $Q$, to the north of $Q_j$, were grown
before $Q_j$.

Therefore, despite Corollary \ref{cor}, the general question remains open:
\begin{openproblem}
  Is there a tile assembly system with less than $2n-1$ tile types,
  that can assemble a filled square at temperature 1, starting from a
  single tile anywhere in the square?
\end{openproblem}

Even though the recent results about its intrinsic universality
\cite{Meunier-2014}, and the present paper, have made significant
advances in that direction, the exact computational power of
temperature~1 systems is still completely unknown. Moreover, the
existence of single tileset simulating, at temperature 1, any other
temperature 1 tile assembly system, is still open. The following open
problem is particularly puzzling, especially in regard of the
impressive results of \cite{Cook-2011}, showing that three-dimensional
temperature 1 tile assembly are capable of Turing computation:

\begin{openproblem}
Is is decidable whether two tile assembly systems have the same terminal assemblies?
\end{openproblem}

Moreover, Theorem \ref{efficient paths} is the first two-dimensional
construction at temperature 1, with less tile types than its
Manhattan diameter, in the general aTAM. On the other hand, our
solution to the original conjecture relies heavily on the fact that
our squares need to be completely filled. This leaves the following
question open:

\begin{openproblem}

For all $n$, is there a tile assembly system
$\mathcal{T}_n=(T_n,\sigma_n,1)$, with $|\dom{\sigma_n}|=1$ and $T_n<2n-1$, whose terminal
assemblies all contain at least a square frame, that is, such that for
all $\alpha\in\termasm{\mathcal{T}_n}$, $(\{0,n-1\}\times\{0,\ldots,n-1\})
\cup (\{0,\ldots,n-1\}\times\{0,n-1\})\subseteq \dom{\alpha}$?

\end{openproblem}

The initial construction of a square with $2n-1$ tile types, by
Rothemund and Winfree, used a fairly simple design. Our result shows
that this is optimal, but it does not discard the possibility of other
designs:

\begin{openproblem}
Is there a way to self-assemble a square of size $n\times n$ at
temperature 1 with $2n-1$ tile types, that is not a trivial variation
of Rothemund and Winfree's ``comb'' design (Figure \ref{comb})?
How many are there?
\end{openproblem}

\section{Acknowledgments}

I thank David Doty and Damien Woods for friendly support and
discussions, Erik Winfree and his group for hosting me during this
research, and Paul Rothemund for encouragements. I am especially
grateful to Damien Woods, for all the comments that helped me
unshuffle a previous, particularly tortured, version of this paper,
and more generally for taking me with him on the journey
\cite{woods2013intrinsic}. Section \ref{introduction} of this paper is
partially taken from \cite{Meunier-2014}, I would like to thank my
co-authors on that paper. Also, Tom Hirschowitz and Christophe
Raffalli suggested the name ``tentacular''.

\bibliographystyle{plain}

\input{carres.bbl}
\appendix
\pagebreak
\section{A printable version of \ref{figure:efficient path 1}}
\label{printable-eff1}
\begin{center}
\includegraphics[scale=42]{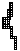}
\end{center}

\pagebreak
\section{A printable version of \ref{figure:efficient path 2}}
\label{printable-eff2}

\begin{center}
\includegraphics[scale=2.1]{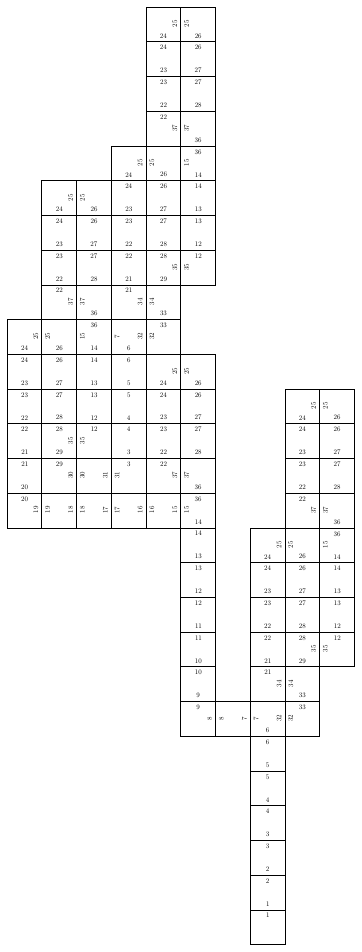}
\end{center}

\end{document}

%% file: intro.tex
Self-assembly is the process through which unorganized, simple,
components automatically coalesce according to simple local rules to
form some kind of target structure. It sounds simple, but the end
result can be extraordinary. For example, researchers have been able
to self-assemble a wide variety of structures experimentally at the
nanoscale, such as regular arrays~\cite{WinLiuWenSee98}, fractal
structures~\cite{RoPaWi04,FujHarParWinMur07}, smiling
faces~\cite{RothOrigami,wei2012complex}, DNA
tweezers~\cite{yurke2000dna}, logic
circuits~\cite{seelig2006enzyme,qian2011scaling}, neural
networks~\cite{qian2011neural}, and molecular
robots\cite{DNARobotNature2010}. These examples are fundamental
because they demonstrate that self-assembly can, in principle, be used
to manufacture specialized geometrical, mechanical and computational
objects at the nanoscale. Potential future applications of nanoscale
self-assembly include the production of smaller, more efficient
microprocessors and medical technologies that are capable of
diagnosing and even treating disease at the cellular level.

Controlling nanoscale self-assembly for the purposes of manufacturing
atomically precise components will require a bottom-up, hands-off
strategy. In other words, the self-assembling units themselves will
have to be ``programmed'' to direct themselves to assemble efficiently
and correctly.  Molecular self-assembly is rapidly becoming a
ubiquitous engineering paradigm, and we need to develop a theory to
inform
us of its algorithmic capabilities and ultimate limitations. %

In 1998, Erik Winfree \cite{Winf98} introduced the abstract Tile
Assembly Model (aTAM), a simplified discrete mathematical model of
algorithmic DNA nanoscale self-assembly pioneered by Seeman
\cite{Seem82}. The aTAM is an asynchronous nondeterministic cellular
automaton that models crystal growth processes. Put another way, the
aTAM essentially augments classical Wang tiling \cite{Wang61} with a
mechanism for sequential growth of a tiling. This contrasts with Wang
tiling in which only the existence of a valid mismatch-free tiling is
considered, and \emph{not} the order of tile placement. In the aTAM,
the fundamental components are translatable but un-rotatable square or
cube \emph{tiles} whose sides are labeled with colored \emph{glues}
colors, each with an integer \emph{strength}.  Two tiles that are
placed next to each other \emph{interact} if the glue colors on their
abutting sides match, and they \emph{bind} if the strengths on their
abutting sides match and sum to at least a certain (integer)
\emph{temperature}. Self-assembly starts from a \emph{seed} tile type
and proceeds nondeterministically and asynchronously as tiles bind to
the seed-containing-assembly. Despite its deliberate simplification,
the aTAM is a computationally expressive model. For example, by using
cooperative binding (that is, by having some of the tiles bind on two
or more sides), Winfree \cite{Winf98} proved that it is Turing
universal, which implies that self-assembly can be directed by a
computer program. Here, we study noncooperative binding.

Tile self-assembly in which tiles can be placed only in a {\em
  noncooperative} fashion is colloquially referred to as
``temperature-1 self-assembly''.  Despite the esoteric name, this is a
fundamental and ubiquitous form of growth: it refers to growth from
{\em growing and branching tips} where each new tile is added if it
can match on at least {\em one side}.

It has been known for some time that a more general form of growth
where some of the tiles must match on two or more sides, i.e.\ {\em
  cooperative} growth, leads to highly non-trivial behavior: arbitrary
Turing machine simulation~\cite{RotWin00, jCCSA}, efficient production
of $n \times n$ squares and other simple shapes using $\Theta(\log
n/\log \log n)$ tile types~\cite{AdChGoHu01}, efficient production of
arbitrary finite connected shapes using a number of tile types that is
within a log factor of the Kolmogorov complexity of the
shape~\cite{SolWin07}, and even intrinsic universality: the existence
of a single tile set that simulates arbitrary tile assembly
systems~\cite{IUSA}.

However, the capabilities of two-dimensional noncooperative
self-assembly remain largely unknown: several generalizations and
restrictions of this model have been studied, that conjectured in all
cases that noncooperative binding could not be as powerful as
cooperative binding \cite{RotWin00, Doty-2011, Reif-2012, Cook-2011,
  Manuch-2010, Patitz-2011}. The first fully general separation
results were only proven recently \cite{Meunier-2014}, in the context
of intrinsic universality~\cite{USA,IUSA,2HAMIU}. However, the
computational capabilities, in the Turing sense, of this model, remain
largely unknown.

The conjecture that we prove in this paper was first stated by
Rothemund and Winfree \cite{RotWin00}\footnote{In Figure 2 of
that paper}: the minimal number of tile types to assemble
$n\times n$ squares is $2n-1$. A restriction of this result, where it
is required that all the tiles be assembled with all their neighbors,
appeared in the same paper.  Moreover, computing the minimal number of
tile types required to \emph{deterministically} assemble a shape from
a seed of size one is known to be NP-complete \cite{ACGHKMR02}, and
$\Sigma_2^p$-complete in the non-deterministic case
\cite{Bryans-2011}.

%% file: tentac.tex
\definecolor{c248ec5}{RGB}{36,142,197}
\definecolor{ccc0000}{RGB}{204,0,0}
\definecolor{c808080}{RGB}{128,128,128}
\begin{tikzpicture}[y=0.80pt, x=0.8pt,yscale=-1, inner sep=0pt, outer sep=0pt]
\begin{scope}[cm={{1.25,0.0,0.0,-1.25,(0.0,157.5)}}]
  \begin{scope}[scale=3.000]
    \path[fill=c248ec5,nonzero rule] (20.0000,42.0000) -- (21.0000,42.0000) --
      (21.0000,41.0000) -- (20.0000,41.0000) -- (20.0000,42.0000) -- cycle;
    \path[fill=c248ec5,nonzero rule] (20.0000,41.0000) -- (21.0000,41.0000) --
      (21.0000,40.0000) -- (20.0000,40.0000) -- (20.0000,41.0000) -- cycle;
    \path[fill=c248ec5,nonzero rule] (20.0000,40.0000) -- (21.0000,40.0000) --
      (21.0000,39.0000) -- (20.0000,39.0000) -- (20.0000,40.0000) -- cycle;
    \path[fill=c248ec5,nonzero rule] (20.0000,39.0000) -- (21.0000,39.0000) --
      (21.0000,38.0000) -- (20.0000,38.0000) -- (20.0000,39.0000) -- cycle;
    \path[fill=c248ec5,nonzero rule] (20.0000,38.0000) -- (21.0000,38.0000) --
      (21.0000,37.0000) -- (20.0000,37.0000) -- (20.0000,38.0000) -- cycle;
    \path[fill=c248ec5,nonzero rule] (20.0000,37.0000) -- (21.0000,37.0000) --
      (21.0000,36.0000) -- (20.0000,36.0000) -- (20.0000,37.0000) -- cycle;
    \path[fill=c248ec5,nonzero rule] (19.0000,37.0000) -- (20.0000,37.0000) --
      (20.0000,36.0000) -- (19.0000,36.0000) -- (19.0000,37.0000) -- cycle;
    \path[fill=c248ec5,nonzero rule] (18.0000,37.0000) -- (19.0000,37.0000) --
      (19.0000,36.0000) -- (18.0000,36.0000) -- (18.0000,37.0000) -- cycle;
    \path[fill=c248ec5,nonzero rule] (17.0000,37.0000) -- (18.0000,37.0000) --
      (18.0000,36.0000) -- (17.0000,36.0000) -- (17.0000,37.0000) -- cycle;
    \path[fill=c248ec5,nonzero rule] (17.0000,38.0000) -- (18.0000,38.0000) --
      (18.0000,37.0000) -- (17.0000,37.0000) -- (17.0000,38.0000) -- cycle;
    \path[fill=c248ec5,nonzero rule] (17.0000,39.0000) -- (18.0000,39.0000) --
      (18.0000,38.0000) -- (17.0000,38.0000) -- (17.0000,39.0000) -- cycle;
    \path[fill=c248ec5,nonzero rule] (16.0000,39.0000) -- (17.0000,39.0000) --
      (17.0000,38.0000) -- (16.0000,38.0000) -- (16.0000,39.0000) -- cycle;
    \path[fill=c248ec5,nonzero rule] (15.0000,39.0000) -- (16.0000,39.0000) --
      (16.0000,38.0000) -- (15.0000,38.0000) -- (15.0000,39.0000) -- cycle;
    \path[fill=c248ec5,nonzero rule] (14.0000,39.0000) -- (15.0000,39.0000) --
      (15.0000,38.0000) -- (14.0000,38.0000) -- (14.0000,39.0000) -- cycle;
    \path[fill=c248ec5,nonzero rule] (13.0000,39.0000) -- (14.0000,39.0000) --
      (14.0000,38.0000) -- (13.0000,38.0000) -- (13.0000,39.0000) -- cycle;
    \path[fill=c248ec5,nonzero rule] (12.0000,39.0000) -- (13.0000,39.0000) --
      (13.0000,38.0000) -- (12.0000,38.0000) -- (12.0000,39.0000) -- cycle;
    \path[fill=c248ec5,nonzero rule] (12.0000,38.0000) -- (13.0000,38.0000) --
      (13.0000,37.0000) -- (12.0000,37.0000) -- (12.0000,38.0000) -- cycle;
    \path[fill=c248ec5,nonzero rule] (12.0000,37.0000) -- (13.0000,37.0000) --
      (13.0000,36.0000) -- (12.0000,36.0000) -- (12.0000,37.0000) -- cycle;
    \path[fill=c248ec5,nonzero rule] (12.0000,36.0000) -- (13.0000,36.0000) --
      (13.0000,35.0000) -- (12.0000,35.0000) -- (12.0000,36.0000) -- cycle;
    \path[fill=c248ec5,nonzero rule] (12.0000,35.0000) -- (13.0000,35.0000) --
      (13.0000,34.0000) -- (12.0000,34.0000) -- (12.0000,35.0000) -- cycle;
    \path[fill=c248ec5,nonzero rule] (12.0000,34.0000) -- (13.0000,34.0000) --
      (13.0000,33.0000) -- (12.0000,33.0000) -- (12.0000,34.0000) -- cycle;
    \path[fill=ccc0000,nonzero rule] (9.0000,33.0000) -- (10.0000,33.0000) --
      (10.0000,32.0000) -- (9.0000,32.0000) -- (9.0000,33.0000) -- cycle;
    \path[fill=c808080,nonzero rule] (10.0000,33.0000) -- (11.0000,33.0000) --
      (11.0000,32.0000) -- (10.0000,32.0000) -- (10.0000,33.0000) -- cycle;
    \path[fill=c808080,nonzero rule] (11.0000,33.0000) -- (12.0000,33.0000) --
      (12.0000,32.0000) -- (11.0000,32.0000) -- (11.0000,33.0000) -- cycle;
    \path[fill=c808080,nonzero rule] (12.0000,33.0000) -- (13.0000,33.0000) --
      (13.0000,32.0000) -- (12.0000,32.0000) -- (12.0000,33.0000) -- cycle;
    \path[fill=c808080,nonzero rule] (12.0000,32.0000) -- (13.0000,32.0000) --
      (13.0000,31.0000) -- (12.0000,31.0000) -- (12.0000,32.0000) -- cycle;
    \path[fill=c808080,nonzero rule] (12.0000,31.0000) -- (13.0000,31.0000) --
      (13.0000,30.0000) -- (12.0000,30.0000) -- (12.0000,31.0000) -- cycle;
    \path[fill=c808080,nonzero rule] (12.0000,30.0000) -- (13.0000,30.0000) --
      (13.0000,29.0000) -- (12.0000,29.0000) -- (12.0000,30.0000) -- cycle;
    \path[fill=c808080,nonzero rule] (12.0000,29.0000) -- (13.0000,29.0000) --
      (13.0000,28.0000) -- (12.0000,28.0000) -- (12.0000,29.0000) -- cycle;
    \path[fill=c808080,nonzero rule] (11.0000,28.0000) -- (12.0000,28.0000) --
      (12.0000,27.0000) -- (11.0000,27.0000) -- (11.0000,28.0000) -- cycle;
    \path[fill=c808080,nonzero rule] (12.0000,28.0000) -- (13.0000,28.0000) --
      (13.0000,27.0000) -- (12.0000,27.0000) -- (12.0000,28.0000) -- cycle;
    \path[fill=c808080,nonzero rule] (10.0000,28.0000) -- (11.0000,28.0000) --
      (11.0000,27.0000) -- (10.0000,27.0000) -- (10.0000,28.0000) -- cycle;
    \path[fill=c808080,nonzero rule] (9.0000,28.0000) -- (10.0000,28.0000) --
      (10.0000,27.0000) -- (9.0000,27.0000) -- (9.0000,28.0000) -- cycle;
    \path[fill=c808080,nonzero rule] (8.0000,28.0000) -- (9.0000,28.0000) --
      (9.0000,27.0000) -- (8.0000,27.0000) -- (8.0000,28.0000) -- cycle;
    \path[fill=c808080,nonzero rule] (8.0000,29.0000) -- (9.0000,29.0000) --
      (9.0000,28.0000) -- (8.0000,28.0000) -- (8.0000,29.0000) -- cycle;
    \path[fill=c808080,nonzero rule] (8.0000,30.0000) -- (9.0000,30.0000) --
      (9.0000,29.0000) -- (8.0000,29.0000) -- (8.0000,30.0000) -- cycle;
    \path[fill=c808080,nonzero rule] (8.0000,31.0000) -- (9.0000,31.0000) --
      (9.0000,30.0000) -- (8.0000,30.0000) -- (8.0000,31.0000) -- cycle;
    \path[fill=c808080,nonzero rule] (8.0000,32.0000) -- (9.0000,32.0000) --
      (9.0000,31.0000) -- (8.0000,31.0000) -- (8.0000,32.0000) -- cycle;
    \path[fill=c808080,nonzero rule] (8.0000,33.0000) -- (9.0000,33.0000) --
      (9.0000,32.0000) -- (8.0000,32.0000) -- (8.0000,33.0000) -- cycle;
    \path[fill=c808080,nonzero rule] (8.0000,34.0000) -- (9.0000,34.0000) --
      (9.0000,33.0000) -- (8.0000,33.0000) -- (8.0000,34.0000) -- cycle;
    \path[fill=c808080,nonzero rule] (8.0000,35.0000) -- (9.0000,35.0000) --
      (9.0000,34.0000) -- (8.0000,34.0000) -- (8.0000,35.0000) -- cycle;
    \path[fill=c808080,nonzero rule] (7.0000,35.0000) -- (8.0000,35.0000) --
      (8.0000,34.0000) -- (7.0000,34.0000) -- (7.0000,35.0000) -- cycle;
    \path[fill=c808080,nonzero rule] (6.0000,35.0000) -- (7.0000,35.0000) --
      (7.0000,34.0000) -- (6.0000,34.0000) -- (6.0000,35.0000) -- cycle;
    \path[fill=c808080,nonzero rule] (5.0000,35.0000) -- (6.0000,35.0000) --
      (6.0000,34.0000) -- (5.0000,34.0000) -- (5.0000,35.0000) -- cycle;
    \path[fill=c808080,nonzero rule] (4.0000,35.0000) -- (5.0000,35.0000) --
      (5.0000,34.0000) -- (4.0000,34.0000) -- (4.0000,35.0000) -- cycle;
    \path[fill=c808080,nonzero rule] (4.0000,34.0000) -- (5.0000,34.0000) --
      (5.0000,33.0000) -- (4.0000,33.0000) -- (4.0000,34.0000) -- cycle;
    \path[fill=c808080,nonzero rule] (4.0000,33.0000) -- (5.0000,33.0000) --
      (5.0000,32.0000) -- (4.0000,32.0000) -- (4.0000,33.0000) -- cycle;
    \path[fill=c808080,nonzero rule] (4.0000,32.0000) -- (5.0000,32.0000) --
      (5.0000,31.0000) -- (4.0000,31.0000) -- (4.0000,32.0000) -- cycle;
    \path[fill=c808080,nonzero rule] (4.0000,31.0000) -- (5.0000,31.0000) --
      (5.0000,30.0000) -- (4.0000,30.0000) -- (4.0000,31.0000) -- cycle;
    \path[fill=c808080,nonzero rule] (4.0000,30.0000) -- (5.0000,30.0000) --
      (5.0000,29.0000) -- (4.0000,29.0000) -- (4.0000,30.0000) -- cycle;
    \path[fill=c808080,nonzero rule] (4.0000,29.0000) -- (5.0000,29.0000) --
      (5.0000,28.0000) -- (4.0000,28.0000) -- (4.0000,29.0000) -- cycle;
    \path[fill=c808080,nonzero rule] (4.0000,28.0000) -- (5.0000,28.0000) --
      (5.0000,27.0000) -- (4.0000,27.0000) -- (4.0000,28.0000) -- cycle;
    \path[fill=c808080,nonzero rule] (4.0000,27.0000) -- (5.0000,27.0000) --
      (5.0000,26.0000) -- (4.0000,26.0000) -- (4.0000,27.0000) -- cycle;
    \path[fill=c808080,nonzero rule] (4.0000,19.0000) -- (5.0000,19.0000) --
      (5.0000,18.0000) -- (4.0000,18.0000) -- (4.0000,19.0000) -- cycle;
    \path[fill=c808080,nonzero rule] (4.0000,18.0000) -- (5.0000,18.0000) --
      (5.0000,17.0000) -- (4.0000,17.0000) -- (4.0000,18.0000) -- cycle;
    \path[fill=c808080,nonzero rule] (4.0000,17.0000) -- (5.0000,17.0000) --
      (5.0000,16.0000) -- (4.0000,16.0000) -- (4.0000,17.0000) -- cycle;
    \path[fill=c808080,nonzero rule] (5.0000,17.0000) -- (6.0000,17.0000) --
      (6.0000,16.0000) -- (5.0000,16.0000) -- (5.0000,17.0000) -- cycle;
    \path[fill=c808080,nonzero rule] (6.0000,17.0000) -- (7.0000,17.0000) --
      (7.0000,16.0000) -- (6.0000,16.0000) -- (6.0000,17.0000) -- cycle;
    \path[fill=c808080,nonzero rule] (7.0000,17.0000) -- (8.0000,17.0000) --
      (8.0000,16.0000) -- (7.0000,16.0000) -- (7.0000,17.0000) -- cycle;
    \path[fill=c808080,nonzero rule] (8.0000,17.0000) -- (9.0000,17.0000) --
      (9.0000,16.0000) -- (8.0000,16.0000) -- (8.0000,17.0000) -- cycle;
    \path[fill=c808080,nonzero rule] (9.0000,17.0000) -- (10.0000,17.0000) --
      (10.0000,16.0000) -- (9.0000,16.0000) -- (9.0000,17.0000) -- cycle;
    \path[fill=c808080,nonzero rule] (10.0000,17.0000) -- (11.0000,17.0000) --
      (11.0000,16.0000) -- (10.0000,16.0000) -- (10.0000,17.0000) -- cycle;
    \path[fill=c808080,nonzero rule] (10.0000,16.0000) -- (11.0000,16.0000) --
      (11.0000,15.0000) -- (10.0000,15.0000) -- (10.0000,16.0000) -- cycle;
    \path[fill=c808080,nonzero rule] (10.0000,15.0000) -- (11.0000,15.0000) --
      (11.0000,14.0000) -- (10.0000,14.0000) -- (10.0000,15.0000) -- cycle;
    \path[fill=c808080,nonzero rule] (10.0000,14.0000) -- (11.0000,14.0000) --
      (11.0000,13.0000) -- (10.0000,13.0000) -- (10.0000,14.0000) -- cycle;
    \path[fill=c808080,nonzero rule] (10.0000,13.0000) -- (11.0000,13.0000) --
      (11.0000,12.0000) -- (10.0000,12.0000) -- (10.0000,13.0000) -- cycle;
    \path[fill=c808080,nonzero rule] (10.0000,12.0000) -- (11.0000,12.0000) --
      (11.0000,11.0000) -- (10.0000,11.0000) -- (10.0000,12.0000) -- cycle;
    \path[fill=c808080,nonzero rule] (9.0000,12.0000) -- (10.0000,12.0000) --
      (10.0000,11.0000) -- (9.0000,11.0000) -- (9.0000,12.0000) -- cycle;
    \path[fill=c808080,nonzero rule] (8.0000,12.0000) -- (9.0000,12.0000) --
      (9.0000,11.0000) -- (8.0000,11.0000) -- (8.0000,12.0000) -- cycle;
    \path[fill=c808080,nonzero rule] (7.0000,12.0000) -- (8.0000,12.0000) --
      (8.0000,11.0000) -- (7.0000,11.0000) -- (7.0000,12.0000) -- cycle;
    \path[fill=c808080,nonzero rule] (6.0000,12.0000) -- (7.0000,12.0000) --
      (7.0000,11.0000) -- (6.0000,11.0000) -- (6.0000,12.0000) -- cycle;
    \path[fill=c808080,nonzero rule] (5.0000,12.0000) -- (6.0000,12.0000) --
      (6.0000,11.0000) -- (5.0000,11.0000) -- (5.0000,12.0000) -- cycle;
    \path[fill=c808080,nonzero rule] (4.0000,12.0000) -- (5.0000,12.0000) --
      (5.0000,11.0000) -- (4.0000,11.0000) -- (4.0000,12.0000) -- cycle;
    \path[fill=c808080,nonzero rule] (3.0000,12.0000) -- (4.0000,12.0000) --
      (4.0000,11.0000) -- (3.0000,11.0000) -- (3.0000,12.0000) -- cycle;
    \path[fill=c808080,nonzero rule] (2.0000,12.0000) -- (3.0000,12.0000) --
      (3.0000,11.0000) -- (2.0000,11.0000) -- (2.0000,12.0000) -- cycle;
    \path[fill=c808080,nonzero rule] (1.0000,12.0000) -- (2.0000,12.0000) --
      (2.0000,11.0000) -- (1.0000,11.0000) -- (1.0000,12.0000) -- cycle;
    \path[fill=c808080,nonzero rule] (0.0000,12.0000) -- (1.0000,12.0000) --
      (1.0000,11.0000) -- (0.0000,11.0000) -- (0.0000,12.0000) -- cycle;
    \path[fill=c808080,nonzero rule] (0.0000,11.0000) -- (1.0000,11.0000) --
      (1.0000,10.0000) -- (0.0000,10.0000) -- (0.0000,11.0000) -- cycle;
    \path[fill=c808080,nonzero rule] (0.0000,10.0000) -- (1.0000,10.0000) --
      (1.0000,9.0000) -- (0.0000,9.0000) -- (0.0000,10.0000) -- cycle;
    \path[fill=c808080,nonzero rule] (0.0000,9.0000) -- (1.0000,9.0000) --
      (1.0000,8.0000) -- (0.0000,8.0000) -- (0.0000,9.0000) -- cycle;
    \path[fill=c808080,nonzero rule] (0.0000,8.0000) -- (1.0000,8.0000) --
      (1.0000,7.0000) -- (0.0000,7.0000) -- (0.0000,8.0000) -- cycle;
    \path[fill=c808080,nonzero rule] (0.0000,7.0000) -- (1.0000,7.0000) --
      (1.0000,6.0000) -- (0.0000,6.0000) -- (0.0000,7.0000) -- cycle;
    \path[fill=c808080,nonzero rule] (0.0000,6.0000) -- (1.0000,6.0000) --
      (1.0000,5.0000) -- (0.0000,5.0000) -- (0.0000,6.0000) -- cycle;
    \path[fill=c808080,nonzero rule] (0.0000,5.0000) -- (1.0000,5.0000) --
      (1.0000,4.0000) -- (0.0000,4.0000) -- (0.0000,5.0000) -- cycle;
    \path[fill=c808080,nonzero rule] (0.0000,4.0000) -- (1.0000,4.0000) --
      (1.0000,3.0000) -- (0.0000,3.0000) -- (0.0000,4.0000) -- cycle;
    \path[fill=c808080,nonzero rule] (0.0000,3.0000) -- (1.0000,3.0000) --
      (1.0000,2.0000) -- (0.0000,2.0000) -- (0.0000,3.0000) -- cycle;
    \path[fill=c808080,nonzero rule] (0.0000,2.0000) -- (1.0000,2.0000) --
      (1.0000,1.0000) -- (0.0000,1.0000) -- (0.0000,2.0000) -- cycle;
    \path[fill=c808080,nonzero rule] (0.0000,1.0000) -- (1.0000,1.0000) --
      (1.0000,0.0000) -- (0.0000,0.0000) -- (0.0000,1.0000) -- cycle;
    \path[fill=c808080,nonzero rule] (4.0000,21.0000) -- (5.0000,21.0000) --
      (5.0000,20.0000) -- (4.0000,20.0000) -- (4.0000,21.0000) -- cycle;
    \path[fill=c808080,nonzero rule] (4.0000,20.0000) -- (5.0000,20.0000) --
      (5.0000,19.0000) -- (4.0000,19.0000) -- (4.0000,20.0000) -- cycle;
    \path[fill=c808080,nonzero rule] (4.0000,24.0000) -- (5.0000,24.0000) --
      (5.0000,23.0000) -- (4.0000,23.0000) -- (4.0000,24.0000) -- cycle;
    \path[fill=c808080,nonzero rule] (4.0000,23.0000) -- (5.0000,23.0000) --
      (5.0000,22.0000) -- (4.0000,22.0000) -- (4.0000,23.0000) -- cycle;
    \path[fill=c808080,nonzero rule] (4.0000,22.0000) -- (5.0000,22.0000) --
      (5.0000,21.0000) -- (4.0000,21.0000) -- (4.0000,22.0000) -- cycle;
    \path[fill=c808080,nonzero rule] (4.0000,26.0000) -- (5.0000,26.0000) --
      (5.0000,25.0000) -- (4.0000,25.0000) -- (4.0000,26.0000) -- cycle;
    \path[fill=c808080,nonzero rule] (4.0000,25.0000) -- (5.0000,25.0000) --
      (5.0000,24.0000) -- (4.0000,24.0000) -- (4.0000,25.0000) -- cycle;
    \path[fill=c248ec5,nonzero rule] (18.0000,26.0000) -- (19.0000,26.0000) --
      (19.0000,25.0000) -- (18.0000,25.0000) -- (18.0000,26.0000) -- cycle;
    \path[fill=c248ec5,nonzero rule] (18.0000,25.0000) -- (19.0000,25.0000) --
      (19.0000,24.0000) -- (18.0000,24.0000) -- (18.0000,25.0000) -- cycle;
    \path[fill=c248ec5,nonzero rule] (18.0000,24.0000) -- (19.0000,24.0000) --
      (19.0000,23.0000) -- (18.0000,23.0000) -- (18.0000,24.0000) -- cycle;
    \path[fill=c248ec5,nonzero rule] (18.0000,23.0000) -- (19.0000,23.0000) --
      (19.0000,22.0000) -- (18.0000,22.0000) -- (18.0000,23.0000) -- cycle;
    \path[fill=c248ec5,nonzero rule] (18.0000,22.0000) -- (19.0000,22.0000) --
      (19.0000,21.0000) -- (18.0000,21.0000) -- (18.0000,22.0000) -- cycle;
    \path[fill=c248ec5,nonzero rule] (18.0000,21.0000) -- (19.0000,21.0000) --
      (19.0000,20.0000) -- (18.0000,20.0000) -- (18.0000,21.0000) -- cycle;
    \path[fill=c248ec5,nonzero rule] (17.0000,21.0000) -- (18.0000,21.0000) --
      (18.0000,20.0000) -- (17.0000,20.0000) -- (17.0000,21.0000) -- cycle;
    \path[fill=c248ec5,nonzero rule] (16.0000,21.0000) -- (17.0000,21.0000) --
      (17.0000,20.0000) -- (16.0000,20.0000) -- (16.0000,21.0000) -- cycle;
    \path[fill=c248ec5,nonzero rule] (15.0000,21.0000) -- (16.0000,21.0000) --
      (16.0000,20.0000) -- (15.0000,20.0000) -- (15.0000,21.0000) -- cycle;
    \path[fill=c248ec5,nonzero rule] (15.0000,22.0000) -- (16.0000,22.0000) --
      (16.0000,21.0000) -- (15.0000,21.0000) -- (15.0000,22.0000) -- cycle;
    \path[fill=c248ec5,nonzero rule] (15.0000,23.0000) -- (16.0000,23.0000) --
      (16.0000,22.0000) -- (15.0000,22.0000) -- (15.0000,23.0000) -- cycle;
    \path[fill=c248ec5,nonzero rule] (14.0000,23.0000) -- (15.0000,23.0000) --
      (15.0000,22.0000) -- (14.0000,22.0000) -- (14.0000,23.0000) -- cycle;
    \path[fill=c248ec5,nonzero rule] (13.0000,23.0000) -- (14.0000,23.0000) --
      (14.0000,22.0000) -- (13.0000,22.0000) -- (13.0000,23.0000) -- cycle;
    \path[fill=c248ec5,nonzero rule] (12.0000,23.0000) -- (13.0000,23.0000) --
      (13.0000,22.0000) -- (12.0000,22.0000) -- (12.0000,23.0000) -- cycle;
    \path[fill=c248ec5,nonzero rule] (11.0000,23.0000) -- (12.0000,23.0000) --
      (12.0000,22.0000) -- (11.0000,22.0000) -- (11.0000,23.0000) -- cycle;
    \path[fill=c248ec5,nonzero rule] (10.0000,23.0000) -- (11.0000,23.0000) --
      (11.0000,22.0000) -- (10.0000,22.0000) -- (10.0000,23.0000) -- cycle;
    \path[fill=c248ec5,nonzero rule] (10.0000,22.0000) -- (11.0000,22.0000) --
      (11.0000,21.0000) -- (10.0000,21.0000) -- (10.0000,22.0000) -- cycle;
    \path[fill=c248ec5,nonzero rule] (10.0000,21.0000) -- (11.0000,21.0000) --
      (11.0000,20.0000) -- (10.0000,20.0000) -- (10.0000,21.0000) -- cycle;
    \path[fill=c248ec5,nonzero rule] (10.0000,20.0000) -- (11.0000,20.0000) --
      (11.0000,19.0000) -- (10.0000,19.0000) -- (10.0000,20.0000) -- cycle;
    \path[fill=c248ec5,nonzero rule] (10.0000,19.0000) -- (11.0000,19.0000) --
      (11.0000,18.0000) -- (10.0000,18.0000) -- (10.0000,19.0000) -- cycle;
    \path[fill=c248ec5,nonzero rule] (10.0000,18.0000) -- (11.0000,18.0000) --
      (11.0000,17.0000) -- (10.0000,17.0000) -- (10.0000,18.0000) -- cycle;
  \end{scope}
  \path[draw=black,->,line join=miter,line cap=butt,miter limit=4.00,line
    width=0.192pt] (37.5000,97.5000) -- (31.5000,49.5000);
  \path[cm={{0.8,0.0,0.0,-0.8,(0.0,42.0)}}] (17.381792,30.399168) node[above
    right] (text4634) {};
  \path[xscale=1.000,yscale=-1.000,fill=black] (35.451103,-41.913467) node[above
    right] (text4692) {$\vect{P_iP_j}$};
\end{scope}
\end{tikzpicture}

%% file: jordan1.tex
\definecolor{cafc6e9}{RGB}{175,198,233}
\definecolor{c808080}{RGB}{128,128,128}
\definecolor{ccc0000}{RGB}{204,0,0}
\begin{tikzpicture}[y=0.80pt, x=0.8pt,yscale=-1, inner sep=0pt, outer sep=0pt]
  \path[shift={(0,0)},fill=cafc6e9,opacity=0.990,even odd rule] (70.0000,93.6878)
    -- (70.0000,57.6878) -- (32.0000,57.6878) -- (32.0000,79.6878) --
    (5.0000,79.6878) -- (5.0000,150.2798) -- (52.3960,150.2798) --
    (52.3960,93.6878) -- (70.0000,93.6878);
\begin{scope}[cm={{1.25,0.0,0.0,-1.25,(0.0,159.97686)}}]
  \path[fill=c808080,nonzero rule] (54.1158,79.9121) -- (60.1286,79.9121) --
    (60.1286,73.8993) -- (54.1158,73.8993) -- (54.1158,79.9121) -- cycle;
  \path[fill=c808080,nonzero rule] (54.1158,73.8993) -- (60.1286,73.8993) --
    (60.1286,67.8864) -- (54.1158,67.8864) -- (54.1158,73.8993) -- cycle;
  \path[fill=c808080,nonzero rule] (54.1158,67.8864) -- (60.1286,67.8864) --
    (60.1286,61.8735) -- (54.1158,61.8735) -- (54.1158,67.8864) -- cycle;
  \path[fill=c808080,nonzero rule] (54.1158,61.8735) -- (60.1286,61.8735) --
    (60.1286,55.8607) -- (54.1158,55.8607) -- (54.1158,61.8735) -- cycle;
  \path[fill=c808080,nonzero rule] (54.1158,55.8607) -- (60.1286,55.8607) --
    (60.1286,49.8478) -- (54.1158,49.8478) -- (54.1158,55.8607) -- cycle;
  \path[fill=c808080,nonzero rule] (30.0643,85.9250) -- (36.0772,85.9250) --
    (36.0772,79.9121) -- (30.0643,79.9121) -- (30.0643,85.9250) -- cycle;
  \path[fill=c808080,nonzero rule] (36.0772,85.9250) -- (42.0901,85.9250) --
    (42.0901,79.9121) -- (36.0772,79.9121) -- (36.0772,85.9250) -- cycle;
  \path[fill=c808080,nonzero rule] (42.0901,85.9250) -- (48.1029,85.9250) --
    (48.1029,79.9121) -- (42.0901,79.9121) -- (42.0901,85.9250) -- cycle;
  \path[fill=c808080,nonzero rule] (48.1029,85.9250) -- (54.1158,85.9250) --
    (54.1158,79.9121) -- (48.1029,79.9121) -- (48.1029,85.9250) -- cycle;
  \path[fill=c808080,nonzero rule] (54.1158,85.9250) -- (60.1286,85.9250) --
    (60.1286,79.9121) -- (54.1158,79.9121) -- (54.1158,85.9250) -- cycle;
  \path[fill=c808080,nonzero rule] (24.0515,85.9250) -- (30.0643,85.9250) --
    (30.0643,79.9121) -- (24.0515,79.9121) -- (24.0515,85.9250) -- cycle;
  \path[fill=c808080,nonzero rule] (24.0515,79.9121) -- (30.0643,79.9121) --
    (30.0643,73.8993) -- (24.0515,73.8993) -- (24.0515,79.9121) -- cycle;
  \path[fill=c808080,nonzero rule] (24.0515,73.8993) -- (30.0643,73.8993) --
    (30.0643,67.8864) -- (24.0515,67.8864) -- (24.0515,73.8993) -- cycle;
  \path[fill=c808080,nonzero rule] (48.1029,55.8607) -- (54.1158,55.8607) --
    (54.1158,49.8478) -- (48.1029,49.8478) -- (48.1029,55.8607) -- cycle;
  \path[fill=c808080,nonzero rule] (42.0901,55.8607) -- (48.1029,55.8607) --
    (48.1029,49.8478) -- (42.0904,49.8481) -- cycle;
  \path[fill=c808080,nonzero rule] (30.0643,55.8607) -- (36.0772,55.8607) --
    (36.0772,49.8478) -- (30.0643,49.8478) -- (30.0643,55.8607) -- cycle;
  \path[fill=c808080,nonzero rule] (24.0515,67.8864) -- (30.0643,67.8864) --
    (30.0643,61.8735) -- (24.0515,61.8735) -- (24.0515,67.8864) -- cycle;
  \path[fill=c808080,nonzero rule] (18.0386,67.8864) -- (24.0515,67.8864) --
    (24.0515,61.8735) -- (18.0386,61.8735) -- (18.0386,67.8864) -- cycle;
  \path[fill=c808080,nonzero rule] (12.0257,67.8864) -- (18.0386,67.8864) --
    (18.0386,61.8735) -- (12.0257,61.8735) -- (12.0257,67.8864) -- cycle;
  \path[fill=c808080,nonzero rule] (6.0129,67.8864) -- (12.0257,67.8864) --
    (12.0257,61.8735) -- (6.0129,61.8735) -- (6.0129,67.8864) -- cycle;
  \path[fill=c808080,nonzero rule] (0.0000,67.8864) -- (6.0129,67.8864) --
    (6.0129,61.8735) -- (0.0000,61.8735) -- (0.0000,67.8864) -- cycle;
  \path[fill=c808080,nonzero rule] (0.0000,61.8735) -- (6.0129,61.8735) --
    (6.0129,55.8607) -- (0.0000,55.8607) -- (0.0000,61.8735) -- cycle;
  \path[fill=c808080,nonzero rule] (0.0000,55.8607) -- (6.0129,55.8607) --
    (6.0129,49.8478) -- (0.0000,49.8478) -- (0.0000,55.8607) -- cycle;
  \path[fill=c808080,nonzero rule] (0.0000,49.8478) -- (6.0129,49.8478) --
    (6.0129,43.8349) -- (0.0000,43.8349) -- (0.0000,49.8478) -- cycle;
  \path[fill=c808080,nonzero rule] (0.0000,43.8349) -- (6.0129,43.8349) --
    (6.0129,37.8221) -- (0.0000,37.8221) -- (0.0000,43.8349) -- cycle;
  \path[fill=c808080,nonzero rule] (0.0000,37.8221) -- (6.0129,37.8221) --
    (6.0129,31.8092) -- (0.0000,31.8092) -- (0.0000,37.8221) -- cycle;
  \path[fill=c808080,nonzero rule] (0.0000,31.8092) -- (6.0129,31.8092) --
    (6.0129,25.7963) -- (0.0000,25.7963) -- (0.0000,31.8092) -- cycle;
  \path[fill=ccc0000,nonzero rule] (36.0772,55.8607) -- (42.0901,55.8607) --
    (42.0901,49.8478) -- (36.0772,49.8478) -- (36.0772,55.8607) -- cycle;
  \path[fill=c808080,nonzero rule] (0.0000,25.7963) -- (6.0129,25.7963) --
    (6.0129,19.7835) -- (0.0000,19.7835) -- (0.0000,25.7963) -- cycle;
  \path[fill=c808080,nonzero rule] (0.0000,19.7835) -- (6.0129,19.7835) --
    (6.0129,13.7706) -- (0.0000,13.7706) -- (0.0000,19.7835) -- cycle;
  \path[fill=c808080,nonzero rule] (0.0000,13.7706) -- (6.0129,13.7706) --
    (6.0129,7.7577) -- (0.0000,7.7577) -- cycle;
  \path[draw=black,dash pattern=on 1.92pt off 0.64pt,line join=miter,line
    cap=butt,miter limit=4.00,dash phase=3.200pt,line width=0.320pt]
    (-5.2039,7.7577) -- (85.2237,7.7577);
  \path[draw=black,dash pattern=on 1.92pt off 0.64pt,line join=miter,line
    cap=butt,miter limit=4.00,dash phase=3.200pt,line width=0.320pt]
    (42.0904,60.8084) -- (42.0904,1.0367);
  \path[cm={{2.88864,0.0,0.0,-2.88864,(22.76365,95.92418)}},fill=black,opacity=0.990,even
    odd rule] (12.7357,-5.6483)arc(-0.062:180.062:0.445)arc(-180.062:0.062:0.445)
    -- cycle;
  \path[xscale=1.000,yscale=-1.000,fill=black] (62.882973,-110.1627) node[above
    right] (text4056) {$A$};
\end{scope}
\end{tikzpicture}

%% file: jordan2.tex
\definecolor{cafc6e9}{RGB}{175,198,233}
\definecolor{cf5f403}{RGB}{245,244,3}
\definecolor{c808080}{RGB}{128,128,128}
\begin{tikzpicture}[y=0.80pt, x=0.8pt,yscale=-1, inner sep=0pt, outer sep=0pt]
\begin{scope}[shift={(6.75483,-42.47672)}]
  \path[shift={(0,0)},fill=cafc6e9,opacity=0.990,even odd rule] (102.6130,52.3291)
    -- (30.0643,52.5706) -- (30.0643,79.6878) -- (5.0000,79.6878) --
    (5.0000,150.2798) -- (102.6130,150.2798) -- (102.6130,52.3291);
\end{scope}
\begin{scope}[cm={{1.25,0.0,0.0,-1.25,(6.75483,117.50014)}}]
  \path[fill=cf5f403,nonzero rule] (54.1158,79.9121) -- (60.1286,79.9121) --
    (60.1286,73.8993) -- (54.1158,73.8993) -- (54.1158,79.9121) -- cycle;
  \path[fill=cf5f403,nonzero rule] (54.1158,73.8993) -- (60.1286,73.8993) --
    (60.1286,67.8864) -- (54.1158,67.8864) -- (54.1158,73.8993) -- cycle;
  \path[fill=cf5f403,nonzero rule] (54.1158,67.8864) -- (60.1286,67.8864) --
    (60.1286,61.8735) -- (54.1158,61.8735) -- (54.1158,67.8864) -- cycle;
  \path[fill=cf5f403,nonzero rule] (54.1158,61.8735) -- (60.1286,61.8735) --
    (60.1286,55.8607) -- (54.1158,55.8607) -- (54.1158,61.8735) -- cycle;
  \path[fill=cf5f403,nonzero rule] (54.1158,55.8607) -- (60.1286,55.8607) --
    (60.1286,49.8478) -- (54.1158,49.8478) -- (54.1158,55.8607) -- cycle;
  \path[fill=c808080,nonzero rule] (30.0643,85.9250) -- (36.0772,85.9250) --
    (36.0772,79.9121) -- (30.0643,79.9121) -- (30.0643,85.9250) -- cycle;
  \path[fill=c808080,nonzero rule] (36.0772,85.9250) -- (42.0901,85.9250) --
    (42.0901,79.9121) -- (36.0772,79.9121) -- (36.0772,85.9250) -- cycle;
  \path[fill=c808080,nonzero rule] (42.0901,85.9250) -- (48.1029,85.9249) --
    (48.1029,79.9121) -- (42.0901,79.9121) -- cycle;
  \path[fill=c808080,nonzero rule] (48.1029,85.9250) -- (54.1158,85.9250) --
    (54.1158,79.9121) -- (48.1029,79.9121) -- (48.1029,85.9250) -- cycle;
  \path[fill=c808080,nonzero rule] (54.1158,85.9250) -- (60.1286,85.9250) --
    (60.1286,79.9121) -- (54.1158,79.9121) -- (54.1158,85.9250) -- cycle;
  \path[fill=c808080,nonzero rule] (24.0515,85.9250) -- (30.0643,85.9250) --
    (30.0643,79.9121) -- (24.0515,79.9121) -- (24.0515,85.9250) -- cycle;
  \path[fill=c808080,nonzero rule] (24.0515,79.9121) -- (30.0643,79.9121) --
    (30.0643,73.8993) -- (24.0515,73.8993) -- (24.0515,79.9121) -- cycle;
  \path[fill=c808080,nonzero rule] (24.0515,73.8993) -- (30.0643,73.8993) --
    (30.0643,67.8864) -- (24.0515,67.8864) -- (24.0515,73.8993) -- cycle;
  \path[fill=cf5f403,nonzero rule] (48.1029,55.8607) -- (54.1158,55.8607) --
    (54.1158,49.8478) -- (48.1029,49.8478) -- (48.1029,55.8607) -- cycle;
  \path[fill=cf5f403,nonzero rule] (42.0901,55.8607) -- (48.1029,55.8607) --
    (48.1029,49.8478) -- (42.0904,49.8481) -- cycle;
  \path[fill=c808080,nonzero rule] (24.0515,67.8864) -- (30.0643,67.8864) --
    (30.0643,61.8735) -- (24.0515,61.8735) -- (24.0515,67.8864) -- cycle;
  \path[fill=c808080,nonzero rule] (18.0386,67.8864) -- (24.0515,67.8864) --
    (24.0515,61.8735) -- (18.0386,61.8735) -- (18.0386,67.8864) -- cycle;
  \path[fill=c808080,nonzero rule] (12.0257,67.8864) -- (18.0386,67.8864) --
    (18.0386,61.8735) -- (12.0257,61.8735) -- (12.0257,67.8864) -- cycle;
  \path[fill=c808080,nonzero rule] (6.0129,67.8864) -- (12.0257,67.8864) --
    (12.0257,61.8735) -- (6.0129,61.8735) -- (6.0129,67.8864) -- cycle;
  \path[fill=c808080,nonzero rule] (0.0000,67.8864) -- (6.0129,67.8864) --
    (6.0129,61.8735) -- (0.0000,61.8735) -- (0.0000,67.8864) -- cycle;
  \path[fill=c808080,nonzero rule] (0.0000,61.8735) -- (6.0129,61.8735) --
    (6.0129,55.8607) -- (0.0000,55.8607) -- (0.0000,61.8735) -- cycle;
  \path[fill=c808080,nonzero rule] (0.0000,55.8607) -- (6.0129,55.8607) --
    (6.0129,49.8478) -- (0.0000,49.8478) -- (0.0000,55.8607) -- cycle;
  \path[fill=c808080,nonzero rule] (0.0000,49.8478) -- (6.0129,49.8478) --
    (6.0129,43.8349) -- (0.0000,43.8349) -- (0.0000,49.8478) -- cycle;
  \path[fill=c808080,nonzero rule] (0.0000,43.8349) -- (6.0129,43.8349) --
    (6.0129,37.8221) -- (0.0000,37.8221) -- (0.0000,43.8349) -- cycle;
  \path[fill=c808080,nonzero rule] (0.0000,37.8221) -- (6.0129,37.8221) --
    (6.0129,31.8092) -- (0.0000,31.8092) -- (0.0000,37.8221) -- cycle;
  \path[fill=c808080,nonzero rule] (0.0000,31.8092) -- (6.0129,31.8092) --
    (6.0129,25.7963) -- (0.0000,25.7963) -- (0.0000,31.8092) -- cycle;
  \path[fill=c808080,nonzero rule] (0.0000,25.7963) -- (6.0129,25.7963) --
    (6.0129,19.7835) -- (0.0000,19.7835) -- (0.0000,25.7963) -- cycle;
  \path[fill=c808080,nonzero rule] (0.0000,19.7835) -- (6.0129,19.7835) --
    (6.0129,13.7706) -- (0.0000,13.7706) -- (0.0000,19.7835) -- cycle;
  \path[fill=c808080,nonzero rule] (0.0000,13.7706) -- (6.0129,13.7706) --
    (6.0129,7.7577) -- (0.0000,7.7577) -- cycle;
  \path[draw=black,dash pattern=on 1.28pt off 0.64pt,line join=miter,line
    cap=butt,miter limit=4.00,line width=0.320pt] (-5.2039,7.7577) --
    (85.2237,7.7577);
  \path[draw=black,dash pattern=on 1.28pt off 0.64pt,line join=miter,line
    cap=butt,miter limit=4.00,line width=0.320pt] (82.0904,93.8001) --
    (82.0904,1.0367);
  \path[draw=black,dash pattern=on 1.28pt off 0.64pt,line join=miter,line
    cap=butt,miter limit=4.00,line width=0.320pt] (9.1831,86.1182) --
    (93.7028,86.1182);
  \path[cm={{0.8,0.0,0.0,-0.8,(-5.40386,94.00011)}},fill=black]
    (77.860168,-5.8974862) node[above right] (text3804) {};
\end{scope}
\end{tikzpicture}

%% file: carres.bbl
\begin{thebibliography}{10}

\bibitem{AdChGoHu01}
Leonard Adleman, Qi~Cheng, Ashish Goel, and Ming-Deh Huang.
\newblock Running time and program size for self-assembled squares.
\newblock In {\em Proceedings of the 33rd Annual ACM Symposium on Theory of
  Computing}, pages 740--748, Hersonissos, Greece, 2001.

\bibitem{ACGHKMR02}
Leonard~M. Adleman, Qi~Cheng, Ashish Goel, Ming-Deh~A. Huang, David Kempe,
  Pablo~Moisset de~Espan\'{e}s, and Paul W.~K. Rothemund.
\newblock Combinatorial optimization problems in self-assembly.
\newblock In {\em Proceedings of the Thirty-Fourth Annual ACM Symposium on
  Theory of Computing}, pages 23--32, 2002.

\bibitem{Bryans-2011}
Nathaniel Bryans, Ehsan Chiniforooshan, David Doty, Lila Kari, and Shinnosuke
  Seki.
\newblock The power of nondeterminism in self-assembly.
\newblock In {\em Proceedings of the Twenty-Second Annual ACM-SIAM Symposium on
  Discrete Algorithms}, SODA '11, pages 590--602. SIAM, 2011.
\newblock Arxiv preprint: \href{http://arxiv.org/abs/1006.2897}{\tt
  arXiv:1006.2897}.

\bibitem{Reif-2012}
Harish Chandran, Nikhil Gopalkrishnan, and John Reif.
\newblock Tile complexity of approximate squares.
\newblock {\em Algorithmica}, 66(1):1--17, 2013.

\bibitem{Cook-2011}
Matthew Cook, Yunhui Fu, and Robert~T. Schweller.
\newblock Temperature 1 self-assembly: deterministic assembly in {3D} and
  probabilistic assembly in {2D}.
\newblock In {\em Proceedings of the 22nd Annual ACM-SIAM Symposium on Discrete
  Algorithms}, pages 570--589, 2011.
\newblock Arxiv preprint: \href{http://arxiv.org/abs/0912.0027}{\tt
  arXiv:0912.0027}.

\bibitem{2HAMIU}
Erik~D. Demaine, Matthew~J. Patitz, Trent~A. Rogers, Robert~T. Schweller,
  Scott~M. Summers, and Damien Woods.
\newblock The two-handed tile assembly model is not intrinsically universal.
\newblock In {\em ICALP: 40th International Colloquium on Automata, Languages
  and Programming}, volume 7965 of {\em LNCS}, pages 400--412, Riga, Latvia,
  July 2013. Springer.
\newblock Arxiv preprint \href{http://arxiv.org/abs/1306.6710}{\tt
  arXiv:1306.6710}.

\bibitem{IUSA}
David Doty, Jack~H. Lutz, Matthew~J. Patitz, Robert~T. Schweller, Scott~M.
  Summers, and Damien Woods.
\newblock The tile assembly model is intrinsically universal.
\newblock In {\em Proceedings of the 53rd Annual IEEE Symposium on Foundations
  of Computer Science}, pages 439--446, October 2012.
\newblock Arxiv preprint: \href{http://arxiv.org/abs/1111.3097}{\tt
  arXiv:1111.3097}.

\bibitem{USA}
David Doty, Jack~H. Lutz, Matthew~J. Patitz, Scott~M. Summers, and Damien
  Woods.
\newblock Intrinsic universality in self-assembly.
\newblock In {\em Proceedings of the 27th International Symposium on
  Theoretical Aspects of Computer Science}, pages 275--286, 2009.
\newblock Arxiv preprint: \href{http://arxiv.org/abs/1001.0208}{\tt
  arXiv:1001.0208}.

\bibitem{Doty-2011}
David Doty, Matthew~J. Patitz, and Scott~M. Summers.
\newblock Limitations of self-assembly at temperature 1.
\newblock {\em Theoretical Computer Science}, 412(1--2):145--158, 2011.
\newblock Arxiv preprint: \href{http://arxiv.org/abs/0906.3251}{\tt
  arXiv:0906.3251}.

\bibitem{FujHarParWinMur07}
Kenichi Fujibayashi, Rizal Hariadi, Sung~Ha Park, Erik Winfree, and Satoshi
  Murata.
\newblock Toward reliable algorithmic self-assembly of {DNA} tiles: A
  fixed-width cellular automaton pattern.
\newblock {\em Nano Letters}, 8(7):1791--1797, 2007.

\bibitem{jCCSA}
James~I. Lathrop, Jack~H. Lutz, Matthew~J. Patitz, and Scott~M. Summers.
\newblock Computability and complexity in self-assembly.
\newblock {\em Theory Comput. Syst.}, 48(3):617--647, 2011.

\bibitem{DNARobotNature2010}
Kyle Lund, Anthony~T. Manzo, Nadine Dabby, Nicole Micholotti, Alexander
  Johnson-Buck, Jeanetter Nangreave, Steven Taylor, Renjun Pei, Milan~N.
  Stojanovic, Nils~G. Walter, Erik Winfree, and Hao Yan.
\newblock Molecular robots guided by prescriptive landscapes.
\newblock {\em Nature}, 465:206--210, 2010.

\bibitem{Manuch-2010}
J\'an Ma\v{n}uch, Ladislav Stacho, and Christine Stoll.
\newblock Two lower bounds for self-assemblies at temperature 1.
\newblock {\em Journal of Computational Biology}, 17(6):841--852, 2010.

\bibitem{Meunier-2014}
Pierre-Etienne Meunier, Matthew~J. Patitz, Scott~M. Summers, Guillaume
  Theyssier, Andrew Winslow, and Damien Woods.
\newblock Intrinsic universality in tile self-assembly requires cooperation.
\newblock 2013.
\newblock Arxiv preprint: \href{http://arxiv.org/abs/1304.1679}{\tt
  arXiv:1304.1679}.

\bibitem{Patitz-2011}
Matthew~J. Patitz, Robert~T. Schweller, and Scott~M. Summers.
\newblock Exact shapes and {T}uring universality at temperature 1 with a single
  negative glue.
\newblock In {\em DNA 17: Proceedings of the Seventeenth International
  Conference on DNA Computing and Molecular Programming}, LNCS, pages 175--189.
  Springer, September 2011.
\newblock Arxiv preprint: \href{http://arxiv.org/abs/1105.1215}{\tt
  arXiv:1105.1215}.

\bibitem{qian2011scaling}
Lulu Qian and Erik Winfree.
\newblock Scaling up digital circuit computation with {DNA} strand displacement
  cascades.
\newblock {\em Science}, 332(6034):1196, 2011.

\bibitem{qian2011neural}
Lulu Qian, Erik Winfree, and Jehoshua Bruck.
\newblock Neural network computation with {DNA} strand displacement cascades.
\newblock {\em Nature}, 475(7356):368--372, 2011.

\bibitem{RothOrigami}
Paul W.~K. Rothemund.
\newblock {Folding {DNA} to create nanoscale shapes and patterns}.
\newblock {\em Nature}, 440(7082):297--302, March 2006.

\bibitem{RotWin00}
Paul W.~K. Rothemund and Erik Winfree.
\newblock The program-size complexity of self-assembled squares (extended
  abstract).
\newblock In {\em STOC '00: Proceedings of the thirty-second annual ACM
  Symposium on Theory of Computing}, pages 459--468, Portland, Oregon, United
  States, 2000. ACM.

\bibitem{RoPaWi04}
Paul~W.K. Rothemund, Nick Papadakis, and Erik Winfree.
\newblock Algorithmic self-assembly of {DNA} {S}ierpinski triangles.
\newblock {\em PLoS Biology}, 2(12):2041--2053, 2004.

\bibitem{seelig2006enzyme}
Georg Seelig, David Soloveichik, David~Yu Zhang, and Erik Winfree.
\newblock Enzyme-free nucleic acid logic circuits.
\newblock {\em science}, 314(5805):1585--1588, 2006.

\bibitem{Seem82}
Nadrian~C. Seeman.
\newblock Nucleic-acid junctions and lattices.
\newblock {\em Journal of Theoretical Biology}, 99:237--247, 1982.

\bibitem{Sipser}
Michael Sipser.
\newblock {\em Introduction to the Theory of Computation}.
\newblock International Thomson Publishing, 1st edition, 1996.

\bibitem{SolWin07}
David Soloveichik and Erik Winfree.
\newblock Complexity of self-assembled shapes.
\newblock {\em SIAM Journal on Computing}, 36(6):1544--1569, 2007.

\bibitem{Wang61}
Hao Wang.
\newblock Proving theorems by pattern recognition -- {II}.
\newblock {\em The Bell System Technical Journal}, XL(1):1--41, 1961.

\bibitem{wei2012complex}
Bryan Wei, Mingjie Dai, and Peng Yin.
\newblock Complex shapes self-assembled from single-stranded {DNA} tiles.
\newblock {\em Nature}, 485(7400):623--626, 2012.

\bibitem{Winf98}
Erik Winfree.
\newblock {\em Algorithmic Self-Assembly of {D}{N}{A}}.
\newblock PhD thesis, California Institute of Technology, June 1998.

\bibitem{WinLiuWenSee98}
Erik Winfree, Furong Liu, Lisa~A. Wenzler, and Nadrian~C. Seeman.
\newblock Design and self-assembly of two-dimensional {D}{N}{A} crystals.
\newblock {\em Nature}, 394(6693):539--44, 1998.

\bibitem{woods2013intrinsic}
Damien Woods.
\newblock Intrinsic universality and the computational power of self-assembly.
\newblock 2013.
\newblock Arxiv preprint:
  \href{http://arxiv.org/abs/1309.1265}{arXiv:1309.1265}.

\bibitem{yurke2000dna}
Bernard Yurke, Andrew~J Turberfield, Allen~P Mills, Friedrich~C Simmel, and
  Jennifer~L Neumann.
\newblock A {DNA}-fuelled molecular machine made of {DNA}.
\newblock {\em Nature}, 406(6796):605--608, 2000.

\end{thebibliography}
